\documentclass[sigconf]{acmart}

\AtBeginDocument{%
  \providecommand\BibTeX{{%
    \normalfont B\kern-0.5em{\scshape i\kern-0.25em b}\kern-0.8em\TeX}}}

\copyrightyear{2021}
\acmYear{2021}
\setcopyright{acmcopyright}\acmConference[EASE 2021]{Evaluation and Assessment in Software Engineering}{June 21--23, 2021}{Trondheim, Norway}
\acmBooktitle{Evaluation and Assessment in Software Engineering (EASE 2021), June 21--23, 2021, Trondheim, Norway}
\acmPrice{15.00}
\acmDOI{10.1145/3463274.3463351}
\acmISBN{978-1-4503-9053-8/21/06}

\begin{document}

\title{HFContractFuzzer: Fuzzing Hyperledger Fabric Smart Contracts for Vulnerability Detection}

\author{Mengjie Ding, Peiru Li, Shanshan Li, He Zhang}
\affiliation{
  \institution{State Key Laboratory of Novel Software Technology, Software Institute, Nanjing University}
  \city{Nanjing}
  \country{China}}
\email{Email: {alliac711@gmail.com, micro.perrylee@outlook.com, lss@nju.edu.cn, hezhang@nju.edu.cn}}






\begin{abstract}
  With its unique advantages such as decentralization and immutability, blockchain technology has been widely used in various fields in recent years. The smart contract running on the blockchain is also playing an increasingly important role in decentralized application scenarios. Therefore, the automatic detection of security vulnerabilities in smart contracts has become an urgent problem in the application of blockchain technology. Hyperledger Fabric is a smart contract platform based on enterprise-level licensed distributed ledger technology. However, the research on the vulnerability detection technology of Hyperledger Fabric smart contracts is still in its infancy. In this paper, we propose HFContractFuzzer, a method based on Fuzzing technology to detect Hyperledger Fabric smart contracts, which combines a Fuzzing tool for golang named go-fuzz and smart contracts written by golang. We use HFContractFuzzer to detect vulnerabilities in five contracts from typical sources and discover that four of them have security vulnerabilities, proving the effectiveness of the proposed method.
\end{abstract}

\ccsdesc[500]{Security and privacy~Vulnerability scanners}

\keywords{Fuzzing; Smart Contract; Security Vulnerability; Hyperledger Fabric; Blockchain}


\maketitle

\renewcommand{\shortauthors}{Ding et al.}
\section{Introduction}
Since 2008, blockchain has been in the public eye. Blockchain is a new application model of computer technology proposed by Satoshi Nakamoto~\cite{nakamoto2019bitcoin}. It not only provides an opportunity for the development of virtual currencies, but also offers new security solutions. Blockchain technology has been widely used in recent years~\cite{lin2017survey,zheng2017overview}, and various explorations of blockchain technology have been emerging, in which the smart contract is an important part. Smart contracts greatly improve the usage scenarios of blockchain and extends blockchain platform from a simple distributed account system to an extremely rich decentralized operating system~\cite{niyuandong2020}. Compared to normal programs, the security issues of smart contracts should be especially paid attention to. On the one hand, smart contracts cannot be changed once deployed. On the other hand, smart contracts generally carry huge economic value. Therefore, the automatic detection of security vulnerabilities in smart contracts has become a problem to be solved in the application of blockchain technology.

The impressive growth of Ethereum and other blockchain platforms based on the Proof-of-Work consensus protocol made evidently the limitations of this approach, which are mostly related to performance~\cite{sousa2018byzantine}. Meanwhile, Hyperledger Fabric(HF) was originally a consortium blockchain led by the International Business Machines Corporation (IBM) and only permitted certain business organizations to participate and share~\cite{ouyang2019}. HF is a system for deploying and operating permissioned blockchains that targets business applications~\cite{androulaki2018hyperledger}. It supports modular consensus protocols, which allows the system to be tailored to particular use cases and trust models~\cite{androulaki2018hyperledger}. This feature allows HF to be customized more effectively to specific business scenarios and brings shorter latency and higher performance.

However, there is not too much research on HF compared to Ethereum at present, mainly because HF came out later than Ethereum, which was created by the Linux Foundation in 2016. In the past two years, HF has received more and more attention in academia. Nevertheless, there are still almost no public papers or tools on detecting vulnerabilities of HF smart contracts. It is worth mentioning that vulnerabilities in HF smart contracts highly differ from that in Ethereum, which is a challenge as well.

Symbolic Execution, Formal Verification, and Fuzzing are the main techniques of smart contract vulnerability detection, and a large amount of research on Ethereum has been conducted using these techniques~\cite{niyuandong2020}. Among them, Fuzzing technology can detect unexpected vulnerabilities, which is more suitable for vulnerability detection of HF smart contracts, because most of its vulnerabilities are beyond our awareness and research at present.

Based on the background above, we propose a HF smart contract Fuzzing method HFContractFuzzer, which is based on Go-fuzz, a Fuzzing tool for golang. Then we use HFContractFuzzer to detect vulnerabilities in five contracts from typical sources and discover that four of them have security vulnerabilities, proving the effectiveness of the method.

The structure of this paper is arranged as follows: Section~\ref{sec:2} systematically summarizes the technical background and related work involved in this research, including HF smart contract, relevant studies about detecting smart contract vulnerabilities, and the Fuzzing technology; Section~\ref{sec:3} presents the specific method HFContractFuzzer of the go-fuzz-based HF smart contract Fuzzing; Section~\ref{sec:4} tests five typical smart contracts to give the feasibility analysis and verification of the HFContractFuzzer method; Section~\ref{sec:5} discusses the contributions and shortcomings of this study; Section~\ref{sec:6} summarizes the content of this paper.

\section{Background and related work}
\label{sec:2}
This section introduces the technical background and related work involved in this paper, including the introductions of HF smart contracts and a research about detecting smart contract vulnerabilities, as well as Fuzzing technology.

\subsection{Hyperledger Fabric Smart Contract}
In 2016, the Linux Foundation launched an open source project to promote the development of blockchain, opening the era of Hyperledger. Hyperledger currently has six technical frameworks. Among them, HF is an enterprise-level licensed distributed ledger technology platform. HF has four functional modules, namely identity management, ledger management, transaction management, and smart contracts~\cite{zhangzengjun2018}. It is responsible for the development of common modules and designs for most industries~\cite{Rub1}, creating an open source ledger managed by the public with an enterprise-level license~\cite{zhangqinghe2018}.

In HF systems and applications, smart contracts are often referred to as chaincodes. Blockchain networks use chaincodes to constrain and regulate the reading or modification of key-value pairs and other state databases in order to accomplish complex business logic. The chaincode has its own execution logic, which performs the logical processing of the account book data according to the custom rules~\cite{parizi2018empirical}. A node supporting endorsement in the network installs and instantiates the chaincode on a node server so that business people can use the client with Fabric-SDK to communicate with the node service and realize the invocation of the chaincode.

\subsection{Detecting smart contract vulnerabilities}
At present, there has been a certain amount of related research work on the technology of Fuzzing and mining smart contract vulnerabilities. Fu et al.~\cite{fumenglin2019} classified and analyzed the security issues of smart contracts in detail, and summarized the test plans of smart contracts. The article pointed out that common smart contract security issues, including reentrancy, permission control, integer overflow, exception handling, short address attacks, manipulation of block timestamps, denial of service attacks, pseudo-random, delegatecall delegation calls, and illegal and transaction issues. In terms of security, Fu et al. proposed four smart contract testing methods: formal verification, Fuzzing, symbol analysis, and taint analysis.

Mythril \cite{Rub5} is a security analysis tool for Ethereum Virtual Machine(EVM) bytecode. It is based on symbol execution and concrete execution. It combines static execution with dynamic execution to improve path coverage and detection accuracy. However, it is difficult to detect some vulnerabilities of deeper hidden or complex test cases. Oyente \cite{luu2016making} is a static symbolic execution tool for EVM bytecode that detects reentrancy, call stack depth, timestamp dependency, and transaction order dependency vulnerabilities, but the static execution suffers from path explosion, resulting in fewer types of coverable vulnerabilities. Securify \cite{tsankov2018securify} is based on pattern matching, which analyzes vulnerabilities given characteristics in smart contracts. Securify works better than Mythril and Oyente, but it doesn’t address digital vulnerabilities. Smartcheck~\cite{tikhomirov2018smartcheck} is an extensible static analysis tool. It can detect arithmetic overflow vulnerability by feature matching, but it can only match at the source code level and does not provide the input data that triggers the vulnerability. MAIAN~\cite{nikolic2018finding} is an automated tool for finding vulnerabilities in blockchain smart contracts that processes the contract's bytecode and attempts to create a series of transactions to find and confirm errors.

Manticore~\cite{mossberg2019manticore} is based on dynamic symbolic execution and is used to analyze smart contracts and binaries. The types of vulnerabilities it can cover include reentrancy, failure to check return values, unprotected functions, timestamp dependencies, tx.origin used for authentication, and integer overflows. Manticore can enumerate the execution state of a contract and verify the security of critical functions. SAFEVM is the first tool to validate low-level EVM code using an existing validation engine developed for C programs, but the accuracy needed to be improved. Ethersplay is an EVM disassembler for analyzing and debugging smart contracts that are compiled or deployed in blockchains. IDA-EVM is an IDA processor module for EVM that allows reverse-engineering analysis of smart contracts without source codes. WANA~\cite{wang2020wana} is a cross-platform smart contract vulnerability detection tool based on web assembly bytecode symbol. It can effectively detect the typical vulnerabilities of Ethereum smart contracts, but it is not possible to conduct large-scale experiments on Ethereum smart contracts to fully assess WANA's accuracy.

Smart contracts are currently centered around Ethereum and EOS. For HF, only Sukrit et al.~\cite{kalra2018zeus} proposed the ZEUS method. ZEUS leverages abstract interpretation and symbolic model checking, as well as the ability to constrain horn clauses to quickly verify secure contracts. It takes a smart contract and a policy specification as input and verifies the smart contract against that policy. ZEUS performs significantly better than Oyente with a low false alarm rate and an order of magnitude improvement in verification time, but it is a formal verification method that detects whether a smart contract violates the policy and cannot detect other vulnerability types.

\subsection{Fuzzing}
Fuzzing is a technique that has been widely used for vulnerability mining in traditional programs. The core idea is to provide a large number of random test cases for the program, monitor the abnormal behavior of the program during the execution process, and explore some of the potential vulnerabilities that are not easy to identify~\cite{bekrar2011finding}. The generation of test cases is one of the key links of Fuzzing, and there are usually two ways, based on generation and based on mutation~\cite{fuyu2019}.

Fuzzing has proven to be very effective in traditional programs. In smart contracts, there are works surveying and categorizing flaws in critical contracts established that Fuzzing using custom user-defined properties might detect up to 63\% of the most severe and exploitable flaws in contracts~\cite{groce2020actual}. At present, there are many Fuzzing methods for smart contracts in Ethereum. Echidna~\cite{Rub4} is an open-source smart contract Fuzzer. It provides a potent out-of-the-box Fuzzing experience and detected, in less than 2 minutes, many reachability targets that required 15 or more minutes with other Fuzzers. ContractFuzzer~\cite{jiang2018contractfuzzer} generates Fuzzing inputs based on the ABI specification for smart contracts and analyzes the smart contract runtime logs to report security vulnerabilities. It found 459 vulnerabilities with high precision in 6,991 smart contracts. In particular, it detected DAO vulnerability that led to a \$60 million loss. EOSFuzzer~\cite{huang2020eosfuzzer} is a general black-box Fuzzing framework, which provides an effective attack and testing method for EOSIO smart contracts.

ILF~\cite{he2019learning} is a new Fuzzing method based on learning to simulate symbolic execution. It uses neural networks to learn Fuzzing strategies from the input data set generated by symbolic execution experts. The ILF is faster, and its code coverage performs better than existing Fuzzers, but it generates things without including the interaction of intelligent contracts. Harvey~\cite{wustholz2020harvey} is an industrial gray box Fuzzer for smart contracts, which uses input forecasting and demand-driven sequence fuzziness to effectively test smart contracts. Harvey has high coverage and vulnerability detection efficiency, it extends the standard grey box Fuzzing prediction method, but it can not be applied into all smart contracts. Confuzzius~\cite{ferreira2021confuzzius} outperforms ILF and Harvey in detection and is a hybrid Fuzzing tester that combines evolutionary Fuzzing and constraint solving. Confuzzius detects far more vulnerabilities than current state-of-the-art Fuzzers and symbol execution tools and significantly fewer false positives, but its constraint-solving approach currently fails to detect the sequence of things in smart contracts. Badger~\cite{noller2018badger} is a hybrid test method for complexity analysis that extends worst-case ambiguity analysis and uses a modified version of the symbolic pathfinder to introduce fuzziness, analysis, and generality, increased coverage, and execution costs, performance is better than Fuzzing and symbolic execution itself. 

However, HF smart contracts have many different characteristics compared with traditional programs and smart contracts in Ethereum, such as programming languages and consensus protocols, which bring challenges to the Fuzzing for smart contracts in HF. In recent years, the majority of scholars on HF smart contracts Fuzzing technology are still in its infancy and it is more challenging to apply Fuzzing technology to HF smart contracts. Simultaneously, the Fuzzing technology also provides opportunities and new ideas for automated detection of HF smart contract vulnerabilities, which is worthy of in-depth study.

\section{HFContractFuzzer}
\label{sec:3}
HF smart contracts can be written in mainstream languages such as Go and Java. HF has provided Go from the early stage of the project with a software development kit, and the Go community has many tools to improve code quality. Therefore, this paper mainly conducts Fuzzing on smart contracts written in Go language and mines the vulnerability information of smart contracts. The compilation and operation of smart contracts rely on the state database in the blockchain network, which are different from traditional programs. Considering these differences, we firstly conduct unit testing of smart contracts to refine the smart contract runtime environment and local call rules. Then we optimize go-fuzz tool. Finally, we propose a smart contract testing method, HFContractFuzzer, to detect vulnerabilities.

\subsection{HFContractFuzzer Basis}
The basis of HFContractFuzzer is HF smart contract unit testing. Unit testing refers to the testing and evaluation of the basic components as well as different modules of the system to be tested \cite{osherove2009art}. Smart contracts rely on the state database of the blockchain network for read-write operation, so direct calls for them cannot be made locally. In this section, two scenarios are used to execute unit testing for smart contracts in order to obtain local call rules for smart contracts. 

\textbf{Developer Mode.}
The developer model requires a simple local network environment to test the smart contract interfaces. This network environment is different from the blockchain network and only supports smart contract operations. The Docker image information required for testing is shown in Table~\ref{table1}. When configuring the environment, we need to set up the fabric-samples package locally, which contains the debug directory. The debug directory contains a sort node, a node, a smart contract container, and a client container. The smart contract container is responsible for running the smart contract under test and the client container is responsible for sending the request messages. The script.sh file of the client container is responsible for creating channels and adding nodes, which can be used to configure the smart contract during testing. The smart contract under test is placed in the chaincode directory, which is mapped to the smart contract container during the testing process.
\begin{table}
\footnotesize
  \caption{Docker Image Information}
   \label{table1}
  \begin{tabular}{ccl}
    \toprule
    REPOSITORY & IMAGE ID & SIZE\\
    \midrule
  hyperledger/fabric-tools & 0f9743ac0662 & 1.49 GB\\
  hyperledger/fabric-tools & 0f9743ac0662 & 1.49 GB\\
  hyperledger/fabric-orderer & 84eaba5388e7 & 120 MB\\
  hyperledger/fabric-peer & 5a52faa5d8c2 & 128 MB\\
  \bottomrule
\end{tabular}
\end{table}

\begin{figure}[h]
  \centering
  \includegraphics[width=\linewidth]{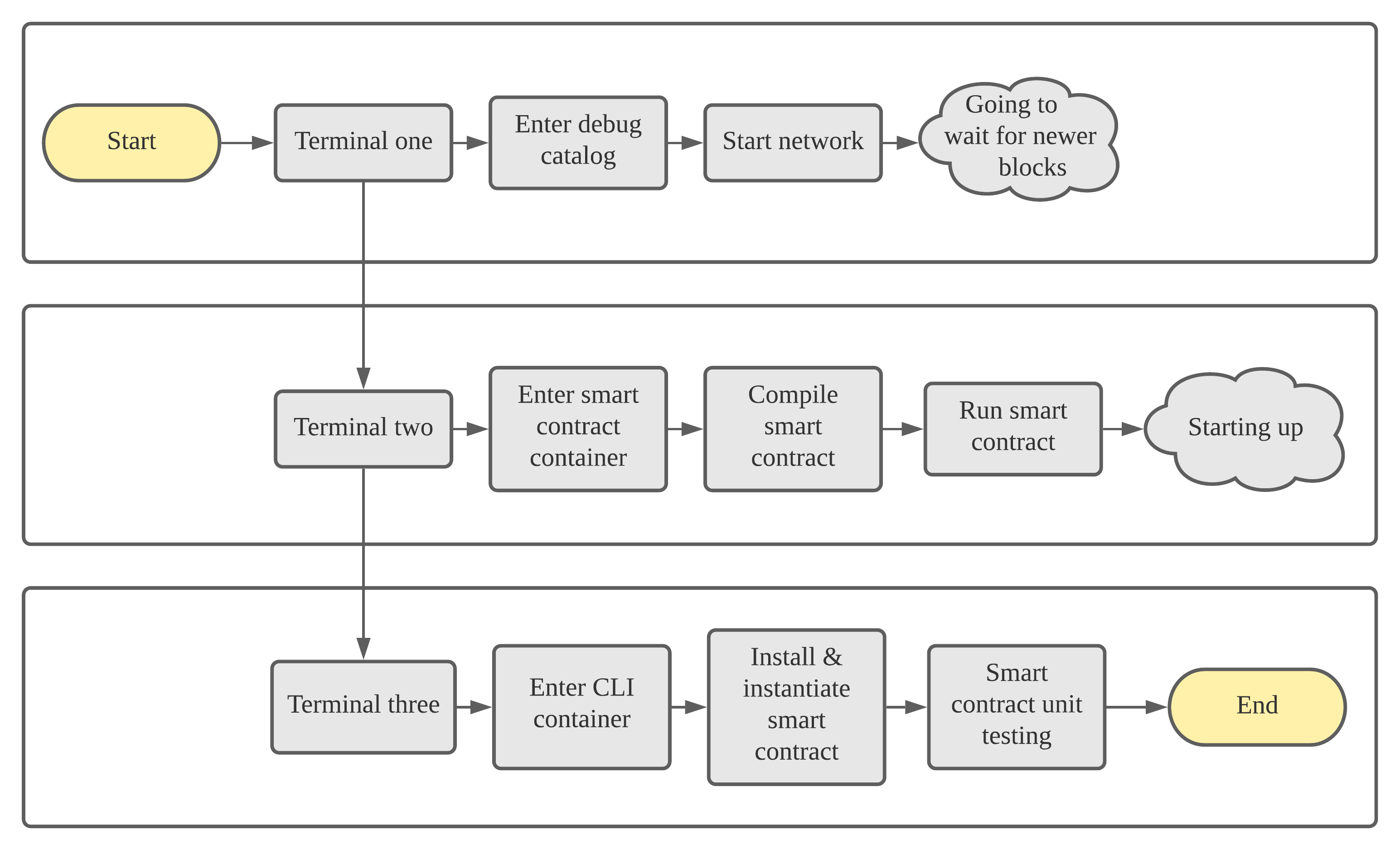}
  \caption{Developer mode unit testing logic flow.}
  \Description{Developer mode unit testing logic flow.}
  \label{figure1}
\end{figure}

The testing process requires starting three terminals, as shown in Figure \ref{figure1}. The first terminal is responsible for starting the network, creating mycc channels, and joining nodes. At this point, the network contains a sorting node server, a node server, a smart contract container, and a client container. The second terminal is responsible for compiling the smart contract and outputting the smart contract's run log. The third terminal is responsible for the configuration of the smart contract. Finally, the third terminal calls the interfaces to execute unit testing for smart contracts.

\textbf{MockStub Class.}
HF provides a MockStub class for unit testing. the MockStub class maintains a map[string][]byte to simulate key-value pairs to complete the work of the state database in the blockchain network. When the PutState and GetState functions are called in a smart contract, it is no longer the state database but the map data in memory. Using MockStub class for unit testing of smart contracts does not require configuration and running network environment, we can complete the smart contract interface calls test locally. The MockStub class provides MockInit and MockInvoke functions to simulate the smart contract calls from endorsement nodes in the blockchain network. Call the Init and Invoke interfaces of smart contracts, the arguments are a string type uuid and a byte array. Init is responsible for the initialization of smart contract, and Invoke is responsible for the selection of smart contract function. Invoke parses byte array, gets function parameters, and calls the different function to complete different Work.

Using developer mode for unit testing of smart contracts simplifies the network environment, but it also involves network configuration and startup. In addition, the developer mode calls the interfaces in the terminal to test the smart contract, and each call requires the execution of a command, specifying the channel name and parameter information, which makes it difficult to realize the input of a large number of random data and is unsuitable for Fuzzing. However, using the MockStub class does not require a blockchain network environment or Docker images. We just execute simple commands to automate the local testing of smart contracts and standardize the smart contract interface function calls. The testing process is the same as the unit testing of traditional Go programs, which has the advantage of simplicity and efficiency. The MockStub class is used to design the rules for calling the smart contract interfaces locally, which facilitates the subsequent testing of smart contracts using Go language testing tools. Therefore, we use MockStub class for exploring the call rules of the smart contract interfaces.

\subsection{Go-fuzz}
At the press conference of the Go v1.5 version, Google announced go-fuzz, testing technology for Go programs. Google completed the Fuzzing with the help of go-fuzz and found 137 errors in the standard library (70 items have been fixed), and 165 other errors~\cite{Rub2}. At present, the most popular go-fuzz tools are Google go-fuzz\footnote{\url{https://github.com/Google/gofuzz}} and Dvyukov go-fuzz\footnote{\url{https://github.com/dvyukov/go-fuzz}}. The Google version is the toolkit officially released by Google. The Dvyukov version is released by the engineer of Dmitry Vyukov. To distinguish, this paper named the two versions of go-fuzz tools G-go-fuzz and D-go-fuzz respectively.

\textbf{G-go-fuzz.} Google officially released a version of go-fuzz~\cite{Rub3}, which uses go-fuzz as a library to fill Go objects with random values. G-go-fuzz supports the generation of multiple types of random data, including single variables, map types, and custom structures.

The G-go-fuzz tool is used to complete the Fuzzing of smart contracts, which is easy to operate and has no special requirements for the test environments. According to the local call rules of the smart contract interface, the random data generation can be completed with the help of go-fuzz library. But G-go-fuzz does not have an exception handling module. When a software vulnerability is found, the terminal reports the exception and suspends the test. This requires testers to suspend testing after discovering a vulnerability, and continue testing after the vulnerability is fixed, which seriously affects the efficiency of testing.

\textbf{D-go-fuzz.} D-go-fuzz is a coverage-guided Fuzzing tool for testing Go packages~\cite{Rub2}. Fuzzing testing is mainly suitable for parsing complex input packages, and strengthening systems that may parse inputs from malicious users.

A D-go-fuzz test unit can be divided into five sections. The corpus is the corpus of Fuzzing, which is responsible for storing the input data. The main.go file in the gen directory is the source of the initial Fuzzing corpus. The fuzz.go file is the key part of the Fuzzing. The Fuzz function uses the random data provided by D-go-fuzz to call the interfaces of the smart contract to complete the test. The crashers directory and suppressions directory are responsible for storing the vulnerability information found during the test. Once D-go-fuzz starts running, it will be an infinite loop genetic algorithm that will constantly generate new test cases based on the initial corpus. D-go-fuzz prints a set of test results at regular intervals. Each group of test results contains a current timestamp and test data information.

Compared with G-go-fuzz, D-go-fuzz has a complicated test process and has certain requirements for the test environment. However, D-go-fuzz includes an exception handling module, which can record vulnerability information when a vulnerability is found during the test, and continue the testing work. In addition, D-go-fuzz can adjust the priority information of the corpus according to the feedback of each test round to improve the efficiency of Fuzzing. So we choose D-go-fuzz for Fuzzing.

\textbf{Optimization of go-fuzz for smart contracts.} D-go-fuzz was originally designed to detect software vulnerabilities in Go language programs, and some of its modules are not applicable to test this particular Go program, smart contracts. Therefore, we optimizes the D-go-fuzz tool in terms of the initial corpus module and the corpus variation process to improve the efficiency of smart contract Fuzzing.

\begin{itemize}
 \item \textbf{Optimization of initial Corpus}. D-go-fuzz is a powerful tool, but the testing process is complicated. The initial corpus data needs to be input manually, which not only is a heavy workload but also brings a high error rate for the system to be tested with a lot of input data. Considering the above situation, the initial corpus generation module of D-go-fuzz is modified as shown in Figure \ref{figure5} to simplify the initial corpus acquisition steps.
 
 \begin{figure}[h]
  \centering
  \includegraphics[width=\linewidth]{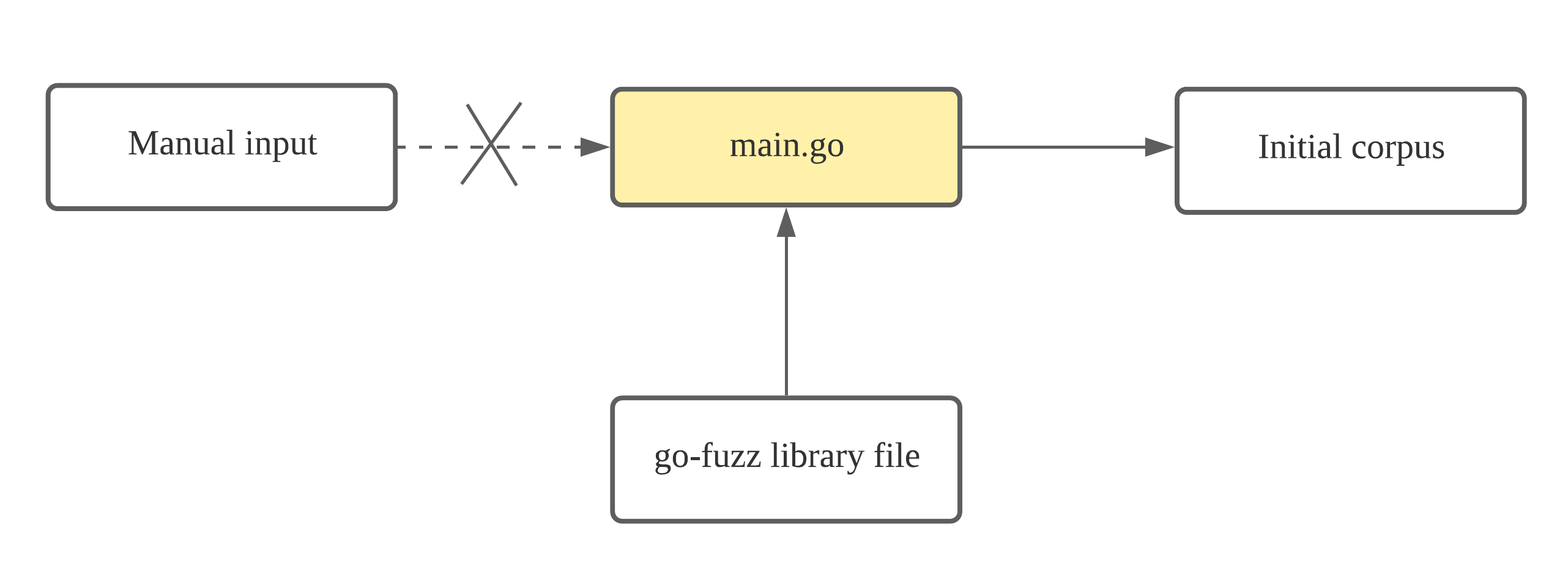}
  \caption{go-fuzz tool improvement diagram.}
  \Description{go-fuzz tool improvement diagram.}
  \label{figure5}
\end{figure}

 The generation of the initial Fuzzing corpus is completed in the main.go file, and data needs to be manually assigned. G-go-fuzz is easy to use, and multiple types of data can be randomly generated by calling go-fuzz library. Use the go-fuzz library in the main.go file to complete the generation of established types of data, and convert it to byte data as the initial corpus for Fuzzing, which can make up for the lack of manual operation and simplify the Fuzzing work.
    \item \textbf{Optimization of mutation process}. After setting the initial corpus, D-go-fuzz adopts a variety of mutation algorithms to mutate the initial corpus. As shown in Table \ref{table2}, there are 20 mutation algorithms. Each mutation algorithm works like its name, such as ID 0, “Remove a range of bytes”, which removes some of the bytes from the initial corpus to generate a new corpus. For smart contracts, some mutation algorithm of D-go-fuzz may have reduced efficiency. In order to improve the efficiency of smart contract Fuzzing, the mutation algorithm of D-go-fuzz is screened and the mutation algorithm with lower efficiency is eliminated.
    
 \begin{table}
 \footnotesize
  \caption{go-fuzz mutation algorithm categories}
  \label{table2}
  \begin{tabular}{ccl}
    \toprule
     ID & Mutation algorithm name\\
    \midrule
  0 & Remove a range of bytes.\\
    1 & Insert a range of random bytes.\\
    2 & Duplicate a range of bytes.\\
    3 & Copy a range of bytes.\\
    4 & Bit flip. Spooky!\\
    5 & Set a byte to a random value.\\
    6 & Swap 2 bytes.\\
    7 & Add/subtract from a byte.\\
    8 & Add/subtract from a uint16.\\
    9 & Add/subtract from a uint32.\\
    10 & Add/subtract from a uint64.\\
    11 & Replace a byte with an interesting value.\\
    12 & Replace an uint16 with an interesting value.\\
    13 & Replace an uint32 with an interesting value.\\
    14 & Replace an ascii digit with another digit.\\
    15 & Replace a multi-byte ASCII number with another number.\\
    16 & Splice another input.\\
    17 & Insert a part of another input.\\
    18 & Insert a literal.\\
    19 & Replace with literal.\\
  \bottomrule
\end{tabular}
\end{table}
 
 We limited the function of randomly selecting mutation algorithms in the fuzzer module of the D-go-fuzz tool, and specified a mutation algorithm each time to perform Fuzzing on the smart contract under test. The test time for each mutation algorithm was five hours. The test report is shown in Table \ref{table4}. The number of Corpus, the number of test execution, the code coverage, and the time of finding exceptions (time-1 is the time of finding the first exception and time-2 is the time of finding the second exception) are selected as the basis of evaluating the mutation algorithms. It can be seen from the test report that the mutation algorithm with ID 1 is less efficient in finding exceptions. The mutation algorithm with ID 19 takes a long time to find anomalies and has too few corpus, so these two mutation algorithms can be eliminated.
 
 \begin{table}
 \footnotesize
  \caption{Mutation algorithm test report}
  \label{table4}
  \begin{tabular}{cccccc}
    \toprule
    ID & corpus & execs & cover  & time-1 & time-2\\
    \midrule
  0 & 261 & 12754433 & 1523 & 1m6s & 3h45m\\
    1 & 409 & 121020447 & 1187 & — & — \\
    2 & 235 & 10739103 & 1509 & 1m9s & 4h0m\\
    3 & 239 & 10712907 & 1515 & 1m6s & 3h49m\\
    4 & 190 & 9927232 & 1510 & 1m9s & 3h59m\\
    5 & 228 & 10772890 & 1527 & 1m6s & 3h49m\\
    6 & 272 & 11075565 & 1521 & 1m6s & 3h39m\\
    7 & 259 & 11434045 & 1521 & 1m6s & 3h47m\\
    8 & 200 & 9652113 & 1490 & 1m6s & 3h53m\\
    9 & 233 & 10809772 & 1527 & 1m6s & 3h44m\\
    10 & 158 & 9195773 & 1457 & 1m9s & 3h52m\\
    11 & 163 & 9088507 & 1475 & 1m3s & 3h58m\\
    12 & 233 & 11434981 & 1518 & 1m9s & 3h48m\\
    13 & 270 & 10573319 & 1526 & 1m6s & 3h1m\\
    14 & 281 & 16859015 & 1519 & 1m15s & 2h39m\\
    15 & 256 & 14502338 & 1526 & 1m6s & 3h9m\\
    16 & 263 & 11850337 & 1522 & 1m9s & 3h13m\\
    17 & 244 & 11610957 & 1527 & 1m9s & 3h44m\\
    18 & 259 & 14245380 & 1522 & 1m6s & 3h41m\\
    19 & 61 & 6958615 & 1359 & 1m3s & 4h30m\\
  \bottomrule
\end{tabular}
\end{table}
 
 \end{itemize}

\subsection{HFContractFuzzer Overview}
Currently, there is no special Fuzzing tool for HF smart contracts. Based on the go-fuzz tool, this section proposes a smart contract Fuzzing method HFContractFuzzer, and the overview diagram of the HFContractFuzzer method is shown in Figure  3. HFContractFuzzer uses the MockStub class to realize the call locally of the smart contract interface and then inputs a lot of random data into smart contracts to detect the potential software vulnerabilities.

 \begin{figure*}[h]
  \centering
  \includegraphics[width=\linewidth]{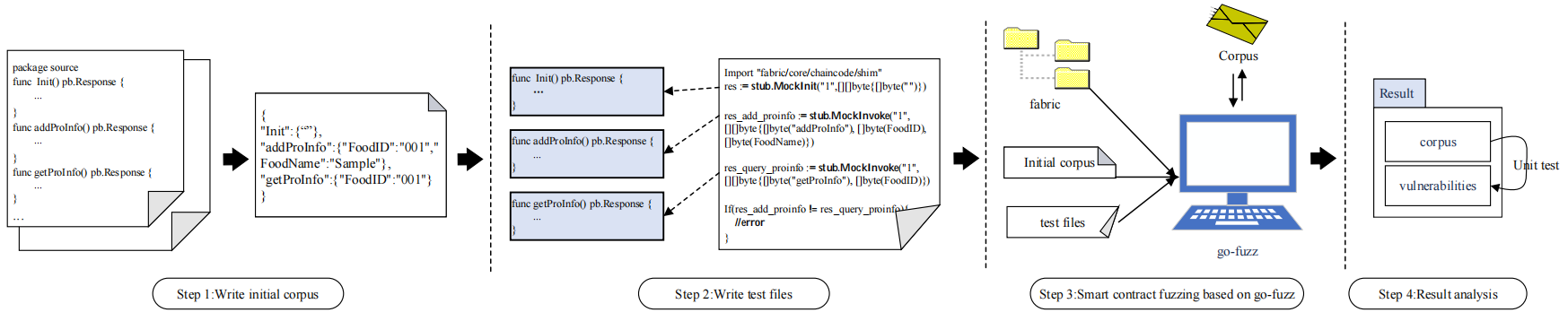}
  \caption{Overview of HFContractFuzzer method.}
  \Description{Overview of HFContractFuzzer method.}
  \label{figure2}
\end{figure*}

\textbf{Write initial corpus.}
First, we provide the unit testing cases of the smart contract to be tested as the initial corpus, Or we use the G-go-fuzz library to complete the generation of data.

\begin{figure}[h]
  \centering
  \includegraphics[width=\linewidth]{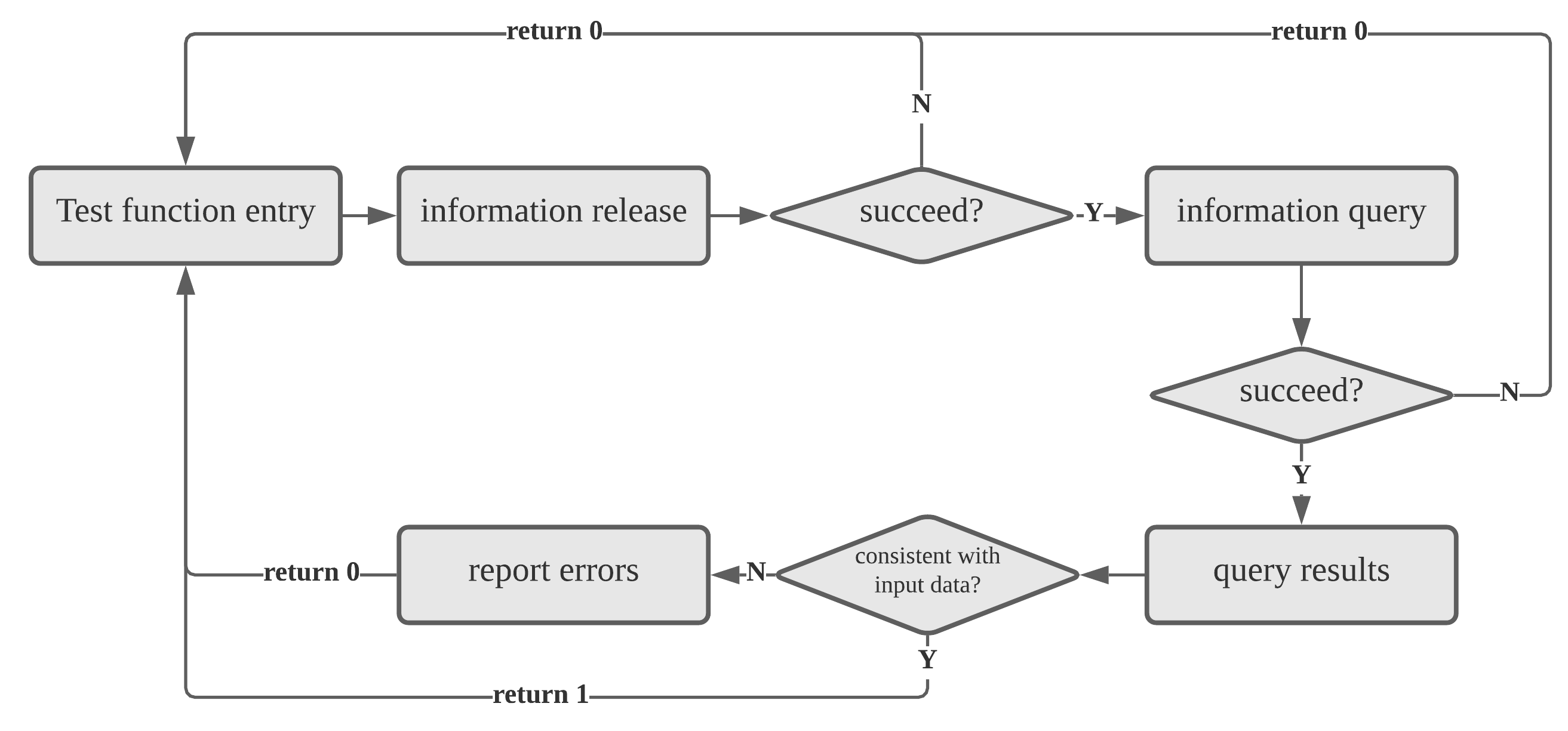}
  \caption{Logical flow of test functions.}
  \Description{Logical flow of test functions.}
  \label{figure3}
  
\end{figure}

\textbf{Write test files.}
We group the release and query for each type of information in the smart contract and write the test functions. Then, we compare the information query results with the published data to detect the logic vulnerabilities of the smart contract. If the test function receives a corpus and returns 0, it indicates that a logical vulnerability may exist, otherwise, it returns 1 to indicate that there are no vulnerabilities, as shown in Figure  \ref{figure3}.

\begin{figure}[h]
  \centering
  \includegraphics[width=\linewidth]{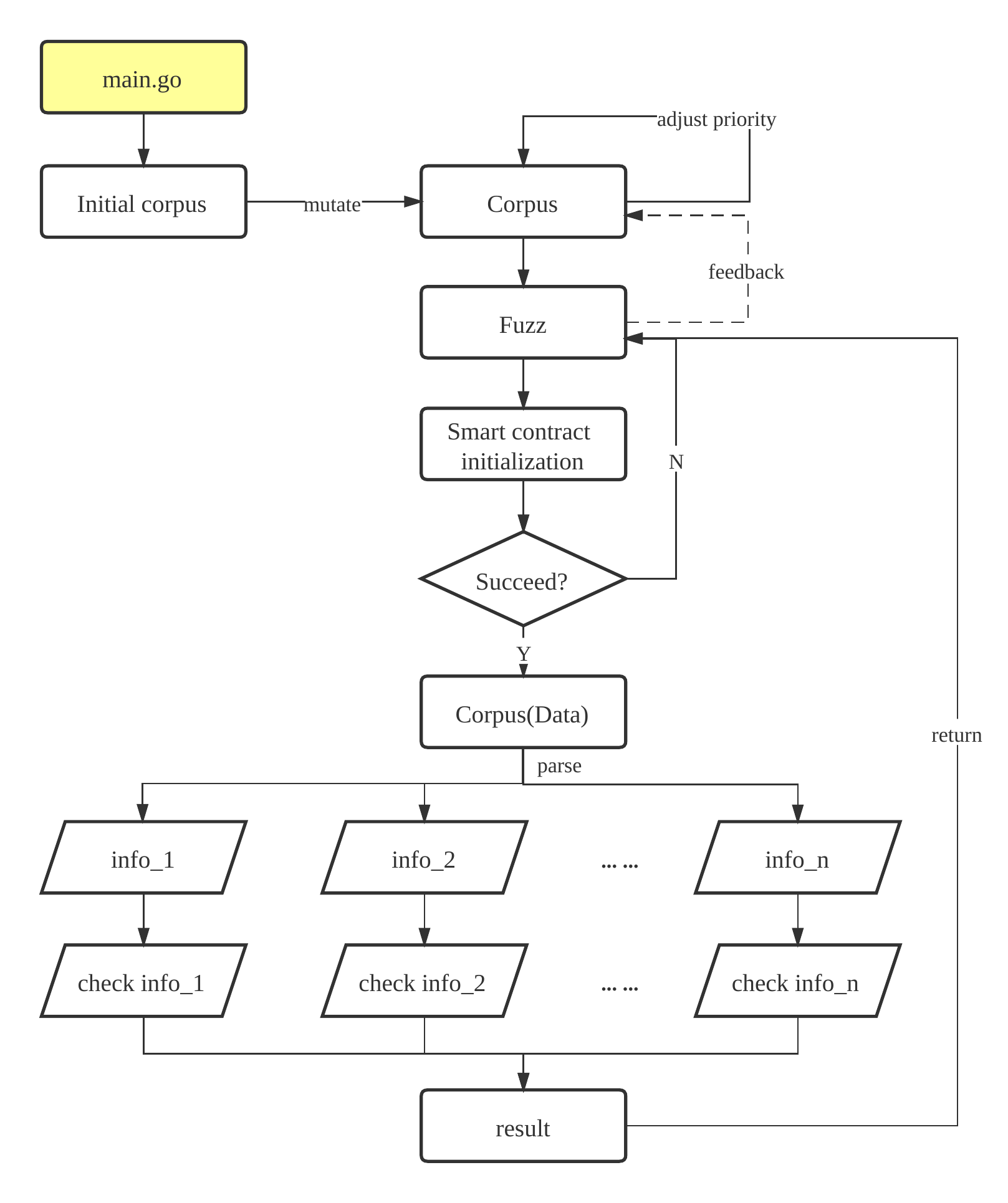}
  \caption{Logical flow of Fuzzing.}
  \Description{Logical flow of Fuzzing.}
  \label{figure4}
\end{figure}

\textbf{Smart contract Fuzzing based on go-fuzz.}
The optimized go-fuzz has a complete exception handling module and high test efficiency, so it is used to fuzz the smart contract. In optimized go-fuzz, the main body of the Fuzzing is implemented in fuzz.go. The Fuzz function is responsible for identifying and parsing the corpus of each test round, transmitting the data to each group of test functions for testing, and combining the results of each group of test functions. The feedback result adjusts the priority of the corpus in the Corpus, as shown in Figure  \ref{figure4}. The exception handling module monitors the entire test execution process, obtains and records exception information.

\textbf{Result analysis.}
The Fuzzing report is displayed on the terminal, and the terminal prints a set of test results at regular intervals. Each group of test results contains a current timestamp and test data information. The corpus is stored in Corpus, and the exception information is stored in crashers. By obtaining the corpus and location information when the exception occurs, and designing unit testing cases, we can determine whether the vulnerability exists and prevent the occurrence of false positives. Finally, we explore the confirmed vulnerabilities, determine the type of vulnerabilities, and analyze the causes of the vulnerabilities.

\begin{table}[!htbp]
\footnotesize
\caption{Unit testing case}
\label{table3}
\begin{tabular}{p{10mm}p{65mm}}
\toprule
 Use Case & Test Data\\
\midrule
 F1 & \{"addProInfo", "001","MiGao", "1", \\ & "2020-01-01", "2020-07-01", "MP01", "FX1234", \\ & "MiGao", "\$10", "Shanxi"\}\\
I1 & \{"addIngInfo", "001", "M1", "NuoMi"\}\\
L1 & \{"addLogInfo", "001", "2020-01-02", "2020-01-02",\\  & "transport", "Shanxi", "Shaanxi",  "MiGao","10 days",\\  &  "Road transport", "China Logistics Company Beijing",  "\$190"\}\\
\bottomrule
\end{tabular}
\end{table}

\begin{table*}[!htbp]
\footnotesize
\caption{Summary of smart contract fuzzy test results}
\label{table5}
\begin{tabular}{p{18mm}p{10mm}p{10mm}p{20mm}p{90mm}}
\toprule
 \textbf{Smart contract} &  \textbf{Source} &  \textbf{Version} & \textbf{Exception type} &  \textbf{Notes}\\
\midrule
example01 & Fabric & 0.6 & — & \\
drm & Caliper & 1.1 & Logic loopholes & It is allowed to write data whose variables are empty strings to the state database,\\  &  &  &  &  but it is considered that this situation does not exist when querying\\
smallbank & Caliper & 1.1 & Integer overflow & The result of adding two integer data greater than 0 is less than 0\\
marbles & IBM & 1.1 & Logic loopholes & In the smart contract, the init function converts the number string to int type, and then\\  &  &  &  & converts the int type to a number string and writes it into the state database. In this\\  &  &  &  & way, the "00" number string meets the data requirements, but it becomes "0" after being\\  &  &  &  &  written into the database\\
\bottomrule
\end{tabular}
\end{table*}

\section{Evaluation}
\label{sec:4}
This section tested five typical smart contracts based on the HF framework to validate the HFContractFuzzer. These smart contracts come from Fabric, Caliper, and IBM including HF version 0.6 and HF version 1.1. At the same time, These smart contracts integrate all the functional modules of HF: Identity Management, account book management, transaction management, smart contracts. And they have the ability to upload and download information. Therefore, they can represent the majority of HF blockchain projects.

Food traceability has long been a hot area of blockchain technology, so we take the food traceability project\footnote{\url{https://github.com/zhazhalaila/hyperledger-simple-app}} among the five smart contracts as an example to give an introduction specifically. The system uses blockchain technology to register and track all kinds of information about food, including basic food information, ingredients, and logistics information. Users can track and query all kinds of information according to the food number to ensure the authenticity of the food.

We tested this smart contract according to the HFContractFuzzer process in Section \ref{sec:3}:

\begin{enumerate}
    \item We used unit testing cases F1, I1, and L1 as the initial corpus for Fuzzing the smart contract under test, as shown in Table \ref{table3}.
     \item We grouped the release and query of basic food information in the smart contract to be tested into one group. Similarly, We grouped ingredient information and logistics information. Then we wrote test files, called the function of each group, and compared the information query results with the published information.
\item We used the optimized go-fuzz tool to test the smart contract. Test time was 25h38min, with execution 233943435 times, cover data  1525, corpus 418, and the test result showed one crasher.

\item We obtained the detailed information of the exception and determine an exception type in the smart contract to be tested.The exception message is "addIngInfo Failed", and the corpus information is "\{"FoodIngInfo":\{"IngID":"$>$"\}\}". According to test results, we designed new test cases based on the corpus and location when the exception occurs, and then performed unit testing, which verifies the type conversion errors. The type conversion error existed in the test file. The input data was a string of "$>$", which was converted to byte type in the smart contract with the help of JSON.Marshal() and stored in the state database. The test file obtained the byte type data in the state database during the query. Used string() to convert to a character string, the converted character string was "\u003e", and the inconsistency with "$>$" caused the system to report an error.

\end{enumerate}

Similarly, we tested other four smart contracts (example01\footnote{\url{https://github.com/hyperledger/fabric/tree/v0.6/examples/chaincode/go/chaincode_example01}},  drm\footnote{\url{https://github.com/caohuilong/Hyperledger-caliper/tree/master/src/contract/fabric}}, smallbank\footnote{\url{https://github.com/caohuilong/Hyperledger-caliper/tree/master/benchmark/smallbank}}, marbles\footnote{\url{https://github.com/IBM-Blockchain/marbles}}) from typical sources, and the results are shown in Table~\ref{table5}. After testing these five typical smart contracts, we discovered that four of them have security vulnerabilities, including three types of vulnerabilities: Type Conversion Errors, Logic loopholes, and Integer overflow, which are also typical and serious in other blockchain platforms such as Ethereum, and then used unit testing to verify that these vulnerabilities do exist. On the one hand, the result proves the feasibility of applying Fuzzing into HF smart contracts to detect unknown vulnerabilities as well as the effectiveness of HFContractFuzzer in vulnerability detection. On the other hand, We find three new vulnerabilities in HF smart contracts that no scholars have discovered before, which contribute to further research on HF smart contracts in the future.

\section{Discussion}
\label{sec:5}
Currently, testing technology for HF smart contracts is still in preliminary exploration stage. Based on this, our paper proposes a Fuzzing method HFContractFuzzer for it, and we use HFContractFuzzer to detect vulnerabilities in five contracts from typical sources. The proposed method discovered that four of them had security vulnerabilities, which proved the effectiveness of it.

Some shortcomings and limitations exist in this work: 

\begin{enumerate}
    \item The go-fuzz tool was originally designed to detect vulnerabilities in Go programs, and there may be performance degradation issues when detecting vulnerabilities of smart contracts. In this paper, although the go-fuzz has been optimized to improve the performance of Fuzzing, the issue still exists and further optimization is needed.
    \item For the evaluation of the proposed Fuzzing method HFContractFuzzer in this paper, we were not able to verify the superiority of it, since we found no detection methods and tools for HF smart contracts available in public papers to support a comparative study. However, we still try our best to show the effectiveness of the proposed method through a carefully designed evaluation process, which selects five typical HF smart contracts, uses our HFContractFuzzer method, and finally detects expected vulnerabilities from all of the selected cases. 
    \item Based on this work, we know three security vulnerabilities exist in HF smart contracts: Type Conversion Errors, Logic loopholes, and Integer overflow. For known vulnerabilities, the effectiveness of other detection methods are is unknown, such as symbolic execution, static analysis, and machine learning. Since these methods have not been incorporated into the HF vulnerability detection domain, the comparison between our method and these promising methods is not within the scope of this study.

\end{enumerate}

\section{Conclusion}
\label{sec:6}
As an important part of the blockchain platform, smart contracts provide rich and credible decentralized application scenarios, and HF provides a platform for smart contracts that are more suitable for business models between enterprises. However, due to the special operating environment and program characteristics of smart contracts, and the predecessors' research foundation on HF smart contracts is less, HF smart contracts are facing a new challenge of security vulnerability detection. Once their vulnerabilities are exploited, it may cause serious economic losses and social impact. This paper proposes the HFContractFuzzer method for HF smart contract detection, combining Fuzzing technology and Go language testing tool go-fuzz. Users provide smart contracts and test cases, HFContractFuzzer automatically generates test results. We use HFContractFuzzer to detect vulnerabilities in five contracts from typical sources and discover four of them have security vulnerabilities, proving the effectiveness of the method. In addition, this paper gives two optimization methods of go-fuzz for smart contracts, initial corpus optimization, and mutation process optimization. With the improvement of vulnerability detection research methods and the maturity of research technology, HF smart contract is expected to be widely used in enterprises and businesses. The research work in this paper is expected to provide useful inspiration and reference for future research on HF smart contract vulnerability detection.

The future work of this study is threefold: 1) Improving the HFContractFuzzer method:  We plan to further optimize the go-fuzz tool and make it more in line with the characteristics of HF smart contracts. 2) Supplementing more evaluation work: We will try to do some more work to evaluate and validate the accuracy and recall of the HFContractFuzzer proposed. 3) Comparison with other methods: We are considering other possible methods that can be adapted for the detection of known HF smart contract vulnerabilities, e,g., symbolic execution, static analysis, and machine learning. After the adaptation and implementation, we will conduct a comparative study to make a comprehensive comparison between the HFContractFuzzer method and other adapted methods.

\begin{acks}
This work is jointly supported by the National Key Research and Development Program of China (No.2019YFE0105500) and the Research Council of Norway (No.309494), as well as the National Natural Science Foundation of China (Grants No.62072227, 61802173), Intergovernmental Bilateral Innovation Project of Jiangsu Province (BZ2020017), and the Innovation
Project of State Key Laboratory for Novel Software Technology (Nanjing University) (ZZKT2021B07).
\end{acks} 
\bibliographystyle{ACM-Reference-Format}
\bibliography{Ref}

\appendix

\end{document}